# Effects of four-wave-mixing in high-power Raman fiber amplifiers

**Wei Liu, Pengfei Ma, Pu Zhou and Zongfu Jiang**

College of Advanced Interdisciplinary Studies, National University of Defense Technology, Changsha, Hunan, 410073, People's Republic of China

E-mail: aiken09@163.com

**Abstract:**
In this work, we derive and present the coupled amplitude equations to describe the evolutions of different spectral components in different transverse modes for Raman fiber amplifiers (RFAs). Both the effects of the four-wave-mixing in the fundamental mode (FM FWM) and the inter-modal four-wave-mixing (IM FWM) on high-power RFAs are demonstrated through numerical simulations. Specifically, effective FM FWM interactions could occur and lead to a drop of the Raman threshold for RFAs by over 50%, despite that the corresponding wave-vector mismatch is rather big. In addition, the IM FWM could also impact the Raman threshold for RFAs with additional generation of the first order Raman Stokes light in the higher-order mode. We also investigate the effects of the intensity fluctuations in the initial inserted pump and seed lasers on high-power RFAs. It reveals that the temporal stability of the initial inserted pump laser have much more significant impacts on high-power RFAs than that of the initial inserted seed laser. Notably, through applying temporal stable laser as the initial inserted pump laser, both the FM FWM and IM FWM effects could be effectively suppressed, and the Raman threshold for high-power RFAs could be increased by over twice.

## 1. Introduction

Raman fiber lasers (RFLs) are attractive light sources which could provide almost arbitrary wavelength in the transmission window of optical fibers [1]. They have been widely investigated and used in optical telecommunication [2] and frequency conversion systems [3]. RFLs are supposed to be the most effective technology to generate wavelength-agile high-power fiber sources [4], and special attention is given to enhance the output powers of RFLs. In recent years, with the development of rare-earth fiber lasers and Raman fiber devices, the output powers of RFLs operating at near- and mid-infrared regions have advanced quickly, which considerably extends the range of RFL applications [5-8]. In particular, kilowatt RFL has been realized at ~1.1 $\mu m$ in 2014 [9], and a 3.89 kW RFL has been reported in 2016 [10], which proves that RFLs have a good perspective for power scaling.

To reduce the power density in the core and further improve the output power, the large-mode area (LMA) fibers are commonly applied in high-power fiber lasers. The LMA fibers support a few transverse modes, which would inevitably introduce the energy conversions among different transverse modes through the inter-modal four-wave mixing (IM FWM) interactions. Previous theoretical and experimental studies have shown that the stimulated Raman scattering (SRS) in optical fibers would be affected by the parametric FWM interactions among the pump, anti-Stokes, and Stokes lights [11-13]. And the relation between the effective Raman gain coefficient and the phase mismatch is given under the undamped pump approximation [12]. Consequently, it could be speculated that the FWM interactions among the pump, first order Raman Stokes and second order Raman Stokes lights would also contribute to the generation of undesired second order Raman Stokes light, and restrict the power scaling of RFLs. However, to best of our knowledge, the impacts of FWM on the power scalability of high-power RFLs, including both the FWM in one transverse mode (mainly the FWM in the fundamental mode, called as FM FWM for short) and the IM FWM, have never been investigated in detail.

In this paper, we aim to clarify the effects of FWM on the power scalability of high-power Raman fiber amplifiers (RFAs). Both the effects of the FM FWM and the IM FWM on high-power RFAs are demonstrated through numerical simulations based on the coupled amplitude equations. This paper is organized as follows. In Section 2, we derive the general coupled amplitude equations for RFAs which takes into account of both the higher-order modes (HOMs) and the second order Raman Stokes light. And the reduced forms of the coupled amplitude equations for high-power case are also given. Solving the coupled amplitude equations numerically, we discuss in Section 3, both the effects of the FM FWM and the IM FWM on high-power RFAs, respectively. In Section 4, we analyze the effects of the intensity fluctuations in the pump and seed lasers on high-power RFAs, and propose the strategies to suppress the two types of FWMs in high-power RFAs. We summarize our results and conclusions in Section 5.

## 2. Theory and numerical modeling

The evolution of the optical field in RFAs could be described through two types of nonlinear propagation equations, i.e., the unified description of generalized nonlinear Schrödinger equation (GNLSE) or the separated description of the coupled amplitude equations. Compared to the GNLSE, the slowly varying envelope is divided into the sum of several slowly varying envelopes with different spectral components in coupled amplitude equations, which could provide a more intuitive approach to describe the interactions among the optical fields with different spectral components. Thus, we apply the coupled amplitude equations to analyze the evolution of the optical field in RFAs here.

*2.1 General coupled amplitude equations*

The general approach to derive the nonlinear propagation equation of an electric field in optical fiber from the Maxwell's equations is outlined in [14], and our starting point is the propagation equation for the envelop of the $k$th mode [15]:

$$\frac{\partial A_k(z,\omega)}{\partial z} = \frac{i\omega}{4}\int F_k^*(x,y) \cdot P_{NL}(r,\omega)e^{-i\beta_k(\omega)z}dxdy \tag{1}$$

where $F$ is the normalized transverse modal profile and the frequency dependence of the modal profile is neglected here; $\beta$ is the propagation constant at the angular frequency $\omega$, $A$ is the slowly varying envelop; $P_{NL}$ is the third-order nonlinear induced polarization accounting for both Kerr and Raman effects, and its Fourier transform could be expressed as:

$$P_{NL}(r,t) = \varepsilon_0 \chi^{(3)} E(r,t) \int R(t-\tau)|E(r,t)|^2 d\tau \tag{2}$$

where $\varepsilon_0$ is the permittivity of vacuum, $\chi^{(3)}$ is the third-order susceptibility. The nonlinear time response function of $R(t)$ could be written as:

$$R(t) = (1-f_R)\delta(t) + f_R h(t) \tag{3}$$

where $f_R$ is the fractional contribution of the Raman response to the total nonlinearity, $\delta(t)$ is the Dirac function and $h(t)$ is the delayed Raman response function. To distinguish the contributions of the different spectral components in the optical field on the nonlinear induced polarization, the total electric field is written as the sum of three continuous wave fields:

$$E(r,t) = \frac{1}{2}\Big[E_p e^{-i\omega_p t} + E_s e^{-i\omega_s t} + E_R e^{-i\omega_R t} + c.c.\Big] \tag{4}$$

where the index $p$, $s$, and $R$ stand for pump light, first order Raman Stokes light, and second order Raman Stokes light, respectively. Each complex amplitude $E_\lambda$ ($\lambda \sim \{p, s, R\}$) could be expressed as a sum of discrete transverse modes:

$$E_\lambda(r,t) = \sum_k F_k(x,y) A_k^{(\lambda)}(z,t) e^{i\beta_k^{(\lambda)}z} \tag{5}$$

After some algebra, we obtain the following coupled amplitude equations for the $k$th mode:

$$\frac{\partial A_k^{(\lambda)}}{\partial z} = -\frac{\alpha_k^{(\lambda)}}{2} - \frac{1}{v_k^{(\lambda)}}\frac{\partial A_k^{(\lambda)}}{\partial t} - \frac{i\beta_{2,k}^{(\lambda)}}{2}\frac{\partial^2 A_k^{(\lambda)}}{\partial t^2}$$

$$+ i\gamma_k^{(\lambda)}(1-f_R) A_k^{(\lambda)} \sum_{l,\lambda'} \left[(2-\delta_{k,l}) f_{kkll} \left|A_l^{(\lambda')}\right|^2\right]$$

$$+ \frac{2i\gamma_k^{(\lambda)} f_R}{3} A_k^{(\lambda)} \cdot h_R \otimes \sum_{l,\lambda'} f_{kkll} \left|A_l^{(\lambda')}\right|^2 \tag{6}$$

$$+ \frac{2i\gamma_k^{(\lambda)} f_R}{3} \sum_{\lambda' \neq \lambda, l} f_{kkll} A_l^{(\lambda')} \cdot \left[h_R e^{i(\omega_\lambda - \omega_{\lambda'})t} \otimes A_k^{(\lambda)} A_l^{(\lambda')*}\right]$$

$$+ \sum f_{klmn} FW_k^{(\lambda)}$$

where $\alpha$ is the loss coefficient, $v$ is the group velocity, $\beta_2$ is the second order dispersion coefficient, $\gamma$ is the nonlinear coefficient; $\otimes$ denotes the convolution operation, $f_{klmn}$ is the nonlinear overlap factor [15].

In the right hand side of (6), the terms in the first line are responsible for the fiber loss and dispersion; the terms in the second line are responsible for the self-phase modulation (SPM) and cross-phase modulation (XPM) effects; the terms in the third line are responsible for the SRS-induced intensity modulation; the terms in the fourth line are responsible for the Raman amplification; the remaining terms in the fifth line are responsible for the frequency combinations of all four waves. It should be noted that most of the FWM terms in (6) are negligible because the occurrence of effective FWM requires the matching of frequencies. Then, the FWM terms could be divided into three categories based on the involved spectral components, i.e., all the four waves in the same spectral component, in two different spectral components, and in three different spectral components. The corresponding frequency matching conditions could be written respectively as:

$$\omega^{(\lambda)} + \omega^{(\lambda)} = \omega^{(\lambda)} + \omega^{(\lambda)} \tag{7}$$

$$\omega^{(\lambda)} + \omega^{(\lambda')} = \omega^{(\lambda)} + \omega^{(\lambda')} \tag{8}$$

$$\omega^{(s)} + \omega^{(s)} = \omega^{(p)} + \omega^{(R)} \tag{9}$$

Special attention is given to the type of FWM terms which satisfies (9), because the corresponding wave-vector mismatch is relatively small compared to the other two categories. And this type of FWM terms could be expressed respectively as:

$$FW_k^{(p)} = i\gamma_k^{(p)}(1-f_R) \sum_{l,m,n} (2-\delta_{l,m}) A_l^{(s)} A_m^{(s)} A_n^{(R)*} e^{-i\Delta_{klmn}^{(p)} z}$$

$$+ \frac{2i\gamma_k^{(p)} f_R}{3} \sum_{l,m,n} A_l^{(s)} \cdot \left[h_R e^{i(\omega_s - \omega_R)t} \otimes A_m^{(s)} A_n^{(R)*}\right] e^{-i\Delta_{klmn}^{(p)} z} \tag{10}$$

$$FW_k^{(s)} = i\gamma_k^{(s)}(1-f_R) \sum_{l,m,n} 2 A_l^{(p)} A_m^{(R)} A_n^{(s)*} e^{i\Delta_{klmn}^{(s)} z}$$

$$+ \frac{2i\gamma_k^{(s)} f_R}{3} \sum_{l,m,n} A_l^{(p)} \cdot \left[h_R e^{i(\omega_R - \omega_s)t} \otimes A_m^{(R)} A_n^{(s)*}\right] e^{i\Delta_{klmn}^{(s)} z} \tag{11}$$

$$+ \frac{2i\gamma_k^{(s)} f_R}{3} \sum_{l,m,n} A_m^{(R)} \cdot \left[h_R e^{i(\omega_p - \omega_s)t} \otimes A_l^{(p)} A_n^{(s)*}\right] e^{i\Delta_{klmn}^{(s)} z}$$

$$FW_k^{(R)} = i\gamma_k^{(R)}(1-f_R) \sum_{l,m,n} (2-\delta_{l,m}) A_l^{(s)} A_m^{(s)} A_n^{(p)*} e^{-i\Delta_{klmn}^{(R)} z}$$

$$+ \frac{2i\gamma_k^{(R)} f_R}{3} \sum_{l,m,n} A_l^{(s)} \cdot \left[h_R e^{i(\omega_s - \omega_p)t} \otimes A_m^{(s)} A_n^{(p)*}\right] e^{-i\Delta_{klmn}^{(R)} z} \tag{12}$$

where the wave-vector mismatch could be expressed as:

$$\Delta_{klmn}^{(p)} = \beta_k^{(p)} + \beta_n^{(R)} - \beta_l^{(s)} - \beta_m^{(s)} \tag{13}$$

$$\Delta_{klmn}^{(s)} = \beta_l^{(p)} + \beta_m^{(R)} - \beta_k^{(s)} - \beta_n^{(s)} \tag{14}$$

$$\Delta_{klmn}^{(R)} = \beta_k^{(R)} + \beta_n^{(p)} - \beta_l^{(s)} - \beta_m^{(s)} \tag{15}$$

*2.2 Reduced forms for the FWM terms in high-power RFAs*

The FWM terms in (10)-(12) could be further classified through two different criterions, i.e., FWM in one transverse mode and IM FWM, instantaneous FWM and delayed Raman-induced FWM. To include and clarify the effects of different types of FWM terms on RFAs, we consider the reduced forms for the FWM terms in high-power RFAs.

In most high-power RFAs, the components of the HOMs are negligible compared to the FM in both the initial inserted seed and pump lasers. Then, the lasers in the FM would be the dominant component all along the fiber in high-power RFAs. Accordingly, the FWM interactions among the HOMs could be negligible compared to that between the FM and the HOMs. Without loss of generality, we consider the representative case when there are two scalar-modes, $LP_{01}$ and $LP_{11}$ modes, in high-power RFAs. Fig. 1 illustrates the schematic diagram for the basic energy conversion processes in the RFA. For simplicity, we call the pump light in $LP_{01}$ mode, the first order Raman Stokes light in $LP_{01}$ mode, the first order Raman Stokes light in $LP_{11}$ mode, the second order Raman Stokes light in $LP_{01}$ mode, and the second order Raman Stokes light in $LP_{11}$ mode as *LP0*, *LS0*, *LS1*, *LR0*, and *LR1*, respectively. As shown in Fig. 1, along with the amplification of *LS0* pumped by *LP0* in the germanium-doped fiber (GDF), the undesired *LS1*, *LR0*, and *LR1* would also occur due to the spontaneous emission noise and the nonlinear effects.

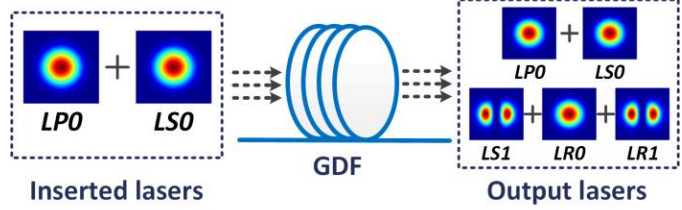

**Fig. 1.** Schematic diagram for basic energy conversion processes in the RFA

Further ignoring the FWM terms in which the wave-vector mismatch would be in the scale of $10^3$ m$^{-1}$, we could focus on two specific types of FWM terms here, i.e. the FM FWM interactions among *LP0*, *LS0*, *LS0* and *LR0*, and the IM FWM interactions among *LP0*, *LS0*, *LS1* and *LR1*. And the reduced form for the FWM terms of *LP0* could be expressed as:

$$FW_0^{(p)} = i\gamma_0^{(p)}(1-f_R) \cdot \begin{Bmatrix} A_0^{(s)} A_0^{(s)} A_0^{(R)*} e^{-i\Delta_1 z} \\ +2A_0^{(s)} A_1^{(s)} A_1^{(R)*} e^{-i\Delta_2 z} \end{Bmatrix}$$
$$+ \frac{2i\gamma_0^{(p)} f_R}{3} \cdot \begin{Bmatrix} A_0^{(s)} \cdot \left[h_R e^{i(\omega_s - \omega_R)t} \otimes A_0^{(s)} A_0^{(R)*}\right] e^{-i\Delta_1 z} \\ +A_0^{(s)} \cdot \left[h_R e^{i(\omega_s - \omega_R)t} \otimes A_1^{(s)} A_1^{(R)*}\right] e^{-i\Delta_2 z} \\ +A_1^{(s)} \cdot \left[h_R e^{i(\omega_s - \omega_R)t} \otimes A_0^{(s)} A_1^{(R)*}\right] e^{-i\Delta_2 z} \end{Bmatrix} \tag{16}$$

where the subscript 0 and 1 stand for $LP_{01}$ and $LP_{11}$ modes, respectively.

In the right hand side of (16), the five terms represent the four typical kinds of FWM terms in high-power RFAs. Specifically, the first term belongs to instantaneous FM FWM; the second term belongs to instantaneous IM FWM; the third term belongs to delayed Raman-induced FM FWM; the fourth and fifth terms belong to delayed Raman-induced IM FWM. The reduced forms for the FWM terms of *LS0*, *LS1*, *LR0*, and *LR1* could be expressed similar to (16), which are shown as follows:

$$FW_0^{(s)} = i\gamma_0^{(s)}(1-f_R) \cdot \begin{Bmatrix} 2A_0^{(p)} A_0^{(R)} A_0^{(s)*} e^{i\Delta_1 z} \\ +2A_0^{(p)} A_1^{(R)} A_1^{(s)*} e^{i\Delta_2 z} \end{Bmatrix} + \frac{2i\gamma_0^{(s)} f_R}{3} \cdot \begin{Bmatrix} A_0^{(R)} \cdot \left[h_R e^{i(\omega_p - \omega_s)t} \otimes A_0^{(p)} A_0^{(s)*}\right] e^{i\Delta_1 z} \\ +A_0^{(p)} \cdot \left[h_R e^{i(\omega_R - \omega_s)t} \otimes A_0^{(R)} A_0^{(s)*}\right] e^{i\Delta_1 z} \\ +A_1^{(R)} \cdot \left[h_R e^{i(\omega_p - \omega_s)t} \otimes A_0^{(p)} A_1^{(s)*}\right] e^{i\Delta_2 z} \\ +A_0^{(p)} \cdot \left[h_R e^{i(\omega_R - \omega_s)t} \otimes A_1^{(R)} A_1^{(s)*}\right] e^{i\Delta_2 z} \end{Bmatrix} \tag{17}$$

$$FW_1^{(s)} = i\gamma_1^{(s)}(1-f_R) \cdot \{2A_0^{(p)}A_1^{(R)}A_0^{(s)*}e^{i\Delta_2 z}\} + \frac{2i\gamma_1^{(s)} f_R}{3} \cdot \begin{Bmatrix} A_1^{(R)} \cdot [h_R e^{i(\omega_p-\omega_s)t} \otimes A_0^{(p)}A_0^{(s)*}]e^{i\Delta_2 z} \\ +A_0^{(p)} \cdot [h_R e^{i(\omega_R-\omega_s)t} \otimes A_1^{(R)}A_0^{(s)*}]e^{i\Delta_2 z} \end{Bmatrix} \quad (18)$$

$$FW_0^{(R)} = i\gamma_0^{(R)}(1-f_R) \cdot \{A_0^{(s)}A_0^{(s)}A_0^{(p)*}e^{-i\Delta_1 z}\} + \frac{2i\gamma_0^{(R)} f_R}{3} \cdot \{A_0^{(s)} \cdot [h_R e^{i(\omega_s-\omega_p)t} \otimes A_0^{(s)}A_0^{(p)*}]e^{-i\Delta_1 z}\} \quad (19)$$

$$FW_1^{(R)} = i\gamma_1^{(R)}(1-f_R) \cdot \{2A_0^{(s)}A_1^{(s)}A_0^{(p)*}e^{-i\Delta_2 z}\} + \frac{2i\gamma_1^{(R)} f_R}{3} \cdot \begin{Bmatrix} A_0^{(s)} \cdot [h_R e^{i(\omega_s-\omega_p)t} \otimes A_1^{(s)}A_0^{(p)*}]e^{-i\Delta_2 z} \\ +A_1^{(s)} \cdot [h_R e^{i(\omega_s-\omega_p)t} \otimes A_0^{(s)}A_0^{(p)*}]e^{-i\Delta_2 z} \end{Bmatrix} \quad (20)$$

We would further analyze and compare the influence of each FWM term on the generation of the second order Raman Stokes light in the following analysis.

*2.3 Generation of the initial inserted lasers and Raman response function*

To apply the coupled amplitude equations to simulate the evolutions of different spectral components in RFAs, we need to construct the initial inserted lasers, including *LP0*, *LS0*, *LS1*, *LR0*, and *LR1*. As for the initial inserted lasers, *LP0* and *LS0*, the typical type of the two lasers is the multi-longitudinal mode fiber laser. Previous studies have shown that there exists intrinsic strong intensity fluctuations in picosecond time scale for typical multi-longitudinal mode fiber lasers [16, 17]. Here we assume that both *LP0* and *LS0* are typical multi-longitudinal Yb-doped mode fiber lasers, and apply the spectral model presented in [18] to construct the two initial inserted lasers. Fig. 2 illustrates the temporal and spectral properties of the constructed initial inserted *LP0* and *LS0*. As shown in Figs. 2(a) and 2(c), there exists strong intensity fluctuations in the temporal evolutions of the two lasers. The corresponding normalized standard deviations (NSDs) for the temporal evolutions of *LP0* and *LS0* are about 0.75 and 0.80, respectively. As shown in Figs. 2(b) and 2(d), the spectral shapes of the two lasers are close to each other. The corresponding root-mean-square (RMS) spectral widths for *LP0* and *LS0* are about 0.32 nm and 0.19 nm, respectively.

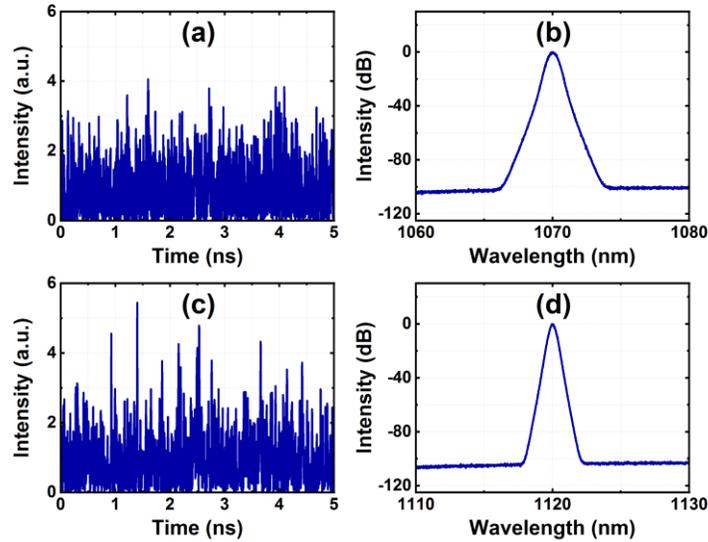

**Fig. 2.** The normalized temporal and spectral properties of the constructed initial inserted LP0 (first line) and LS0 (second line)

As for the initial inserted LS1, *LR0*, and *LR1*, they generally origin from the spontaneous emission noise in RFAs. Accordingly, the spontaneous emission noise term should be added into the coupled amplitude equations (6). Refer to the theory of spontaneous emission noise in open resonators [19], we consider that the real and imaginary parts of spontaneous emission noise satisfy the properties of Gaussian white noise, and its intensity is proportional to the gain coefficient. Accordingly, it can be analyzed as Gaussian stochastic process with zero mean value satisfies:

$$\begin{cases} \langle N_k^{(\lambda)}(z,\omega)N_k^{(\lambda)*}(z',\omega')\rangle = 2D_k^{(\lambda)}\delta(z-z')\delta(\omega-\omega') \\ 2D_k^{(\lambda)} = \dfrac{\hbar\omega^3}{\pi c^2}n(\omega)\cdot g_k^{(\lambda)}(z,\omega)\cdot n_{sp} \end{cases} \quad (21)$$

where $n_{sp}=1/(\exp(\hbar(\omega+\omega_0)/k_BT)-1)$, represents the average mode occupation number in equilibrium; $k_B$ is the Boltzmann constant; $T$ is the environmental temperature; $g$ is the Raman gain coefficient.

To give a more accurate description of the Raman response function, an extended Raman response function is applied in following expressions [20]:

$$h_R(t) = \sum_{k=1}^{13}\frac{b_k}{\omega_k}\exp(\eta_k t)\exp(-\Gamma_k^2 t^2/4)\sin(\omega_k t)\Theta(t) \quad (22)$$

where the values of the coefficients $b_k$, $\omega_k$, $\eta_k$ and $\Gamma_k$ are cited by the reference [20], and $\Theta(t)$ is the Heaviside step function.

## 3. Simulation results

Based on the above model, we simulate the evolutions of different spectral components in a typical high-power RFA with commercial double-clad passive fiber. The core diameter is 20 $\mu m$ and the inner cladding diameter is 400 $\mu m$ (NA=0.75). This fiber supports four scalar modes, $LP_{01}$, $LP_{02}$, $LP_{11}$ and $LP_{21}$ modes, and we focus on the interactions between the $LP_{01}$ and $LP_{11}$ mode here. When setting the nonlinear overlap factor for the FM FWM as 1, then the calculated nonlinear overlap factor for the IM FWM is about 0.64. Besides, the wave-vector mismatches $\Delta_1$ (FM FWM) and $\Delta_2$ (IM FWM) are calculated to be about 72.9 $m^{-1}$ and 1.8 $m^{-1}$, respectively. The other major simulation parameters are shown in Table I. For simplicity, the loss coefficient is set to be the same at different wavelength.

TABLE I

| Parameter | Value | Parameter | Value |
|---|---|---|---|
| $\lambda_p$ | 1070 nm | $\lambda_s$ | 1120 nm |
| $\lambda_R$ | 1175 nm | $v_p$ | $2.068\times10^8$ m/s |
| $v_s$ | $2.069\times10^8$ m/s | $v_R$ | $2.07\times10^8$ m/s |
| $\beta_{2p}$ | 15.6 ps$^2$/km | $\beta_{2s}$ | 11.8 ps$^2$/km |
| $\beta_{2R}$ | 7.4 ps$^2$/km | $\gamma_p$ | 0.5 W$^{-1}$/km |
| $\gamma_s$ | 0.48 W$^{-1}$/km | $\gamma_R$ | 0.46 W$^{-1}$/km |
| $g_p$ | $5.79\times10^{-14}$ m/W | $g_s$ | $5.53\times10^{-14}$ m/W |
| $g_R$ | $5.27\times10^{-14}$ m/W | $\alpha$ | 0.002 m$^{-1}$ |
| $L$ | 20 m | $P_s$ | 50 W |

*3.1 Basic properties of the RFA*

We first demonstrate the basic properties of the RFA at a pump power of 1.5 kW. Figs. 3(a) and 3(b) illustrate the power distribution of different spectral components along the fiber in the RFA. As shown in Fig. 3(a), the output powers of *LP0* and *LS0* are about 0.26 kW and 1.16 kW, respectively. Thus, most of the power of *LP0* is converted into the power of *LS0*, and the corresponding power conversion efficiency is about 77%. As shown in Fig. 3(b), the powers of *LS1*, *LR0* and *LR1* would increase along the fiber, and the corresponding output powers are about -32.2 dBm, 37.6 dBm (5.8 W) and -20.7 dBm, respectively. Thus, the output power of *LR0* is much higher than that of *LR1* despite that they all origin from the spontaneous emission noise. The corresponding power ratio of *LR0* is about 0.5%, and the power ratios of *LS1* and *LR1* could be negligible (<-60 dB) at the output port. The power ratios of *LS1*, *LR0*, and *LR1* are defined by their powers verse the power of *LS0* here and in the following analysis. Figs. 3(c) and 3(d) illustrate the temporal and spectral properties of the output *LS0* in the RFA. As shown in Fig. 3(c), there also exists strong intensity fluctuations in the temporal evolution of the output *LS0*, which is similar to that of the initial inserted pump and seed lasers shown in Figs. 2(a) and 2(c). The NSD for the temporal evolution of the output *LS0* is about 0.82, which is a little bigger than that of the initial inserted *LP0* and *LS0*. Compared to the spectral property of the inserted *LS0* shown in Fig. 2(d), obvious spectral broadening of the output *LS0* is observed in Fig. 3(d). The RMS spectral width of the output *LS0* is about 2.4 nm, which corresponds to a spectral broadening factor of as much as 12.6.

An important parameter for the power scalability of a RFA is the threshold of the second order Raman Stokes light (called as Raman threshold for short). When a RFA reaches its Raman threshold, the power of *LS0* would convert into the power of *LR0* or *LR1* quickly and restrict further power scaling of the RFA. To obtain the Raman threshold for the RFA, we simulate the properties of the RFA at different pump powers.

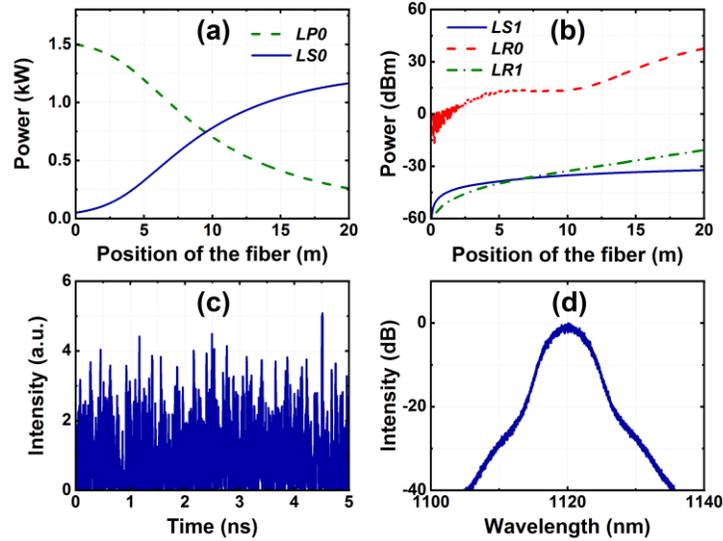

**Fig. 3.** The power distributions, the temporal and spectral properties of the output LS0 in the RFA: (a) the power distributions of LP0 and LS0; (b) the power distributions of LS1, LR0, and LR1; (c) the normalized temporal evolution; (d) the normalized spectral intensity

Fig. 4 illustrates the power ratios of the output *LR0* and *LR1* at different pump powers. As shown in Fig. 4, the power ratio of the output *LR0* increases quickly along with the pump power, while the power ratio of the output *LR1* increases slightly along with the pump power. In addition, the power ratio of the output *LR0* would exceed -20 dB at a pump power of about 1.6 kW. Defining the Raman threshold for the RFA as the output power of *LS0* when the power ratio of the output *LR0* or *LR1* reaches -20 dB, then the Raman threshold for the RFA is about 1.31 kW, which is limited by the occurrence of *LR0* here.

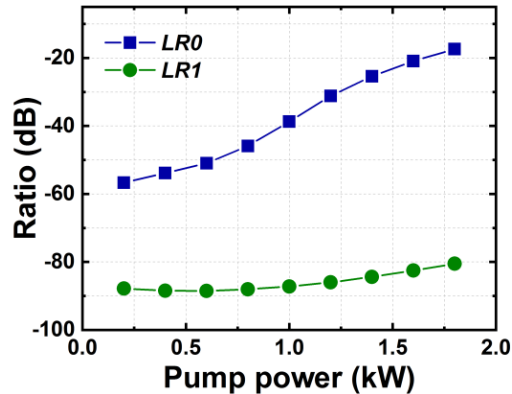

**Fig. 4.** The power ratios of the output LR0 and LR1 at different pump powers

*3.2 Effects of the FM FWM on the RFA*

To clarify the contribution of the FM FWM to the amplification of *LR0* in the RFA, we conduct contrast simulations by omitting the FWM terms in (6). Fig. 5 illustrates the power ratios of the output *LR0* and *LR1* at different pump powers when ignoring the FWM effects in the RFA. As shown in Fig. 5, the power ratio of the output *LR0* still increases more quickly than that of the output *LR1* along with the pump power. And the Raman threshold for the RFA is calculated to be about 2.71 kW, which is also limited by the occurrence of *LR0*.

Combined with the results shown in Fig. 4, the FM FWM leads to a drop of the Raman threshold for the RFA by over 50%.

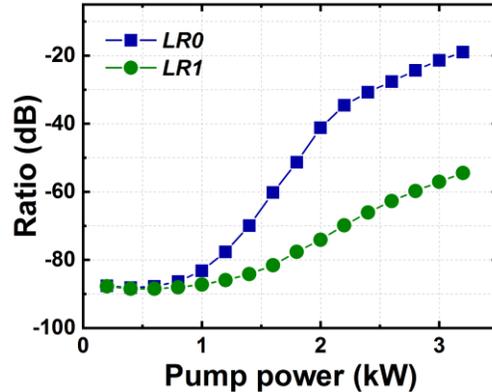

**Fig. 5.** The power ratios of the output LR0 and LR1 at different pump powers when ignoring the FWM effects in the RFA

To further distinguish the impacts of the instantaneous FM FWM and the delayed Raman-induced FM FWM on the RFA, we obtain the Raman thresholds for the RFA when only considering the instantaneous FM FWM terms or the delayed Raman-induced FM FWM terms in (6), respectively. The corresponding Raman thresholds for the RFA are calculated to be about 1.26 kW and 1.57 kW. Thus, the effects of the instantaneous FM FWM on the RFA are stronger than that of the delayed Raman-induced FM FWM.

To this end, we have shown that the FM FWM in normal dispersion region would contribute to the amplification of *LR0*, and lead to obvious decrease of the Raman threshold for the RFA. In other word, effective FM FWM interactions occur in the RFA despite that the wave-vector mismatch is rather big. The mechanism behind this phenomenon is similar to the gain-induced phase-matching [21]. Specifically, when there exist considerable power variations for *LP0* and *LS0* within one coherence length, the total contribution of the FM FWM term in one coherence length would be nonzero. As a consequence, *LR0* never goes back to its initial value, but is progressively constructed as it propagates along the fiber.

*3.3 Effects of the IM FWM on the RFA*

In the section A, the Raman threshold for the RFA is limited by the occurrence of *LR0*. And it could be inferred that the effects of the IM FWM on the RFA are negligible despite that the wave-vector mismatch of the IM FWM is much smaller than that of FM FWM. The main reason for this result is that the power of *LS1* along the fiber is much lower than that of *LS0* (as shown in Fig. 3). When *LS1* only origins from the spontaneous emission noise, and the power ratio of *LS1* keeps below -60 dB along the fiber at different pump powers. However, in practical high-power RFAs, the fiber fusion and the perturbations of the refractive index might lead to the additional generation of *LS1*. Accordingly, the practical power ratios of *LS1* along the fiber would be much larger than the case when *LS1* only origins from the spontaneous emission noise. The impacts of the perturbations of the refractive index on the generation of *LS1* could be analyzed by incorporating a mode coupling coefficient in (6). To distinguish the effects of the IM FWM induced mode coupling from the refractive index change induced mode coupling, we simplify the influence of the additional generation of *LS1* by assuming a small proportion of *LS1* in the initial inserted seed.

To stress the effects of the IM FWM on the RFA, we further ignore the FM FWM terms in the following simulations. Fig. 6 illustrates the power ratios of the output *LR1* at different pump powers when the power proportions of *LS1* in the initial inserted seed are $10^{-6}$, $10^{-5}$, $10^{-4}$, $10^{-3}$, and $10^{-2}$, respectively. Here, the power proportion of *LS1* is defined by the power of *LS1* verse the power of *LS0* in the initial inserted seed. As shown in Fig. 6, the power ratios of the output *LR1* increase along with the pump power for all the five curves, and the corresponding Raman thresholds for the five cases are about 2.26 kW, 1.87 kW, 1.49 kW, 1.12 kW, and 0.83 kW, respectively. Compared to the Raman threshold for the RFA when ignoring the FWM effects (about 2.71 kW), the IM FWM leads to a drop of the Raman threshold by about 17%, 31%, 45%, 59%, and 69% for the five cases, respectively. We may notice that the Raman threshold for the RFA would be lower than 1.31 kW, when the power proportions of *LS1* in the initial inserted seed is over $10^{-3}$. In other word, the Raman threshold for the RFA could be limited by occurrence of *LR1* rather than *LR0*. Therefore, the effects of IM FWM on the RFA are closely related to the additional generation of *LS1*, and it could also impact the Raman threshold for RFAs with additional generation of *LS1*.

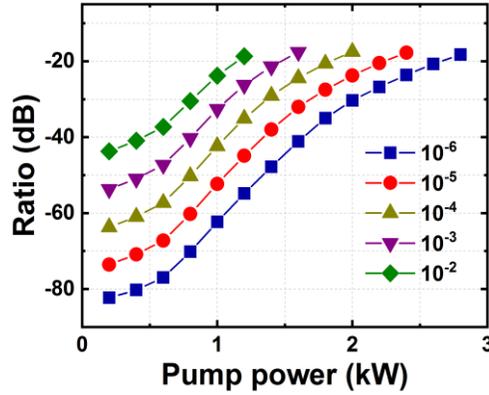

**Fig. 6.** The power ratios of the output LR1 at different pump powers when the power proportions of LS1 in the initial inserted seed are different

## 4. Control of the FWMs in RFAs

To further scale the output power of RFAs, we need to suppress the FWM effects in RFAs. As for the IM FWM, it is clear that the IM FWM could be weakened by suppressing the additional generation of *LS1* in RFAs. Accordingly, the common strategies to suppress generation of *LS1* in fiber amplifiers, such as increasing the relative loss of the HOMs, would also be beneficial to weaken the IM FWM in RFAs. However, as for the FM FWM, it could be speculated that increasing wave-vector mismatch might not be effective to suppress the FM FWM in RFAs.

Previous studies have shown that the temporal properties of the pump and the seed lasers would impact the temporal and spectral properties of RFAs [22, 23]. Thus, it could be speculated that the strong intensity fluctuations in *LP0* and *LS0* might also impact the FWM interactions in RFAs. In the above sections, we have illustrated the properties of the RFA when the initial inserted *LP0* and *LS0* are typical multi-longitudinal mode fiber lasers which induces strong intensity fluctuations in picosecond time scale. In this section, we would consider the contrary cases when the initial inserted *LP0* and *LS0* are replaced by temporal stable lasers, such as phase-modulated single-frequency fiber lasers [24], respectively.

To investigate the impacts of the intensity fluctuations in the pump and the seed lasers on the FM FWM interactions, we first consider the case when *LS1* only origins from the spontaneous emission noise. Fig. 7 illustrates the temporal and spectral properties of the output *LS0* at a pump power of 1.5 kW, when the initial inserted *LS0* or *LP0* is replaced by the temporal stable laser. As shown in Figs. 7(a) and 7(b), when applying temporal stable laser as the initial inserted *LS0*, the temporal and spectral properties of the output *LS0* are similar to the original case shown in Figs. 3(c) and 3(d). The corresponding NSD for the temporal evolution and the RMS spectral width are 0.74 and 2.23 nm, respectively, which are also close to the original case. However, when applying temporal stable laser as the initial inserted *LP0*, the properties of the output *LS0* are quite different from the original case. As shown in Fig. 7(c), the output *LS0* could still keep relatively stable, despite that there exists strong intensity fluctuations in the inserted *LS0* shown in Fig. 2(c). The corresponding NSD for the temporal evolution is about 0.29, which is much smaller than that of the original case (about 0.82). As shown in Fig. 7(d), the spectral broadening phenomenon is much weaker than that the original case. The RMS spectral width is only about 0.27 nm, which corresponds to a spectral broadening factor of about 1.4.

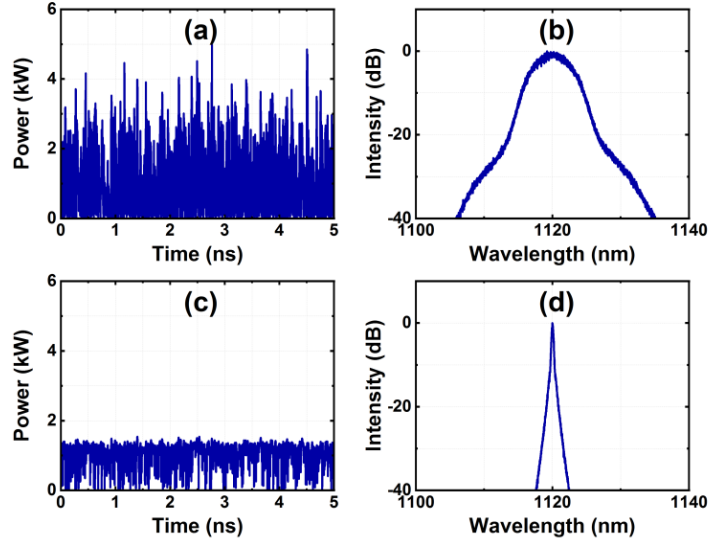

**Fig. 7.** The temporal and spectral properties of the output LS0 at a pump power of 1.5 kW, when the initial inserted LS0 (first line) or LP0 (second line) is the temporal stable laser

To further clarify the impacts of the intensity fluctuations in *LP0* and *LS0* on the FM FWM interactions, we simulate the corresponding power ratios of the output *LR0* at different pump powers. Fig. 8 illustrates the power ratios of the output *LR0* at different pump powers for the two cases. As shown in Fig. 8, when applying temporal stable laser as the initial inserted *LS0*, the growth trend for the curve is similar to the original case. However, the power ratio of the output *LR0* grows much slower along with the pump power, when applying temporal stable laser as the initial inserted *LP0*. The Raman thresholds for the RFA are about 1.68 kW and 4.55 kW, which corresponds to an increase of the Raman threshold by about 28% and 247% for the two cases, respectively. Thus, the FM FWM interactions in the RFA could be effectively suppressed by applying the temporal stable laser as the initial inserted pump laser.

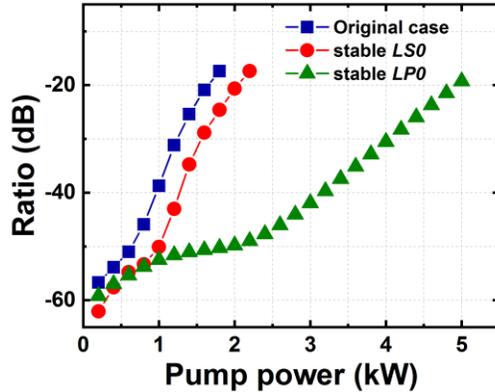

**Fig. 8.** The power ratios of the output LR0 at different pump powers when the initial inserted LS0 or LP0 is the temporal stable laser.

To investigate the impacts of the intensity fluctuations in the pump and the seed lasers on the IM FWM interactions, we consider the case when the power proportion of *LS1* in the initial inserted seed is $10^{-2}$. Fig. 9 illustrates the power ratios of the output *LR1* at different pump powers when the initial inserted *LS0* or *LP0* is replaced by the temporal stable laser, respectively. As shown in Fig. 9, when applying temporal stable laser as the initial inserted *LS0*, the growth trend for the curve is similar to the original case. However, the power ratio of the output *LR1* grows much slower along with the pump power, when applying temporal stable laser as the initial inserted *LP0*. The Raman thresholds for the RFA are about 1.01 kW and 3.07 kW for the two cases, which corresponds to an increase of the Raman threshold by about 22% and 270%, respectively. Thus, the IM FWM interactions in the RFA could be effectively suppressed by applying the temporal stable laser as the initial inserted pump laser.

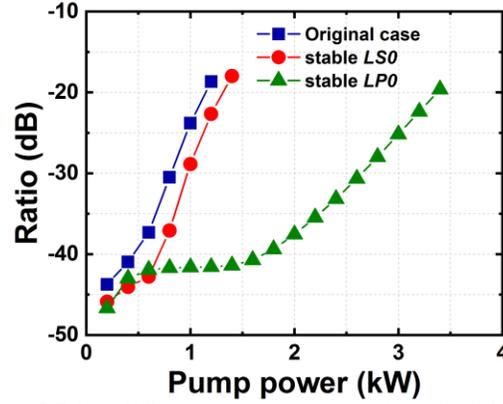

**Fig. 9.** The power ratios of the output LR1 at different pump powers when the initial inserted LS0 or LP0 is the temporal stable laser.

We have also calculated the Raman threshold for the RFA when both the initial inserted *LP0* and *LS0* are temporal stable lasers. The corresponding Raman thresholds for the RFA are about 4.94 kW and 3.18 kW, when *LS1* only origins from the spontaneous emission noise and the power proportion of *LS1* in the initial inserted seed is $10^{-2}$, respectively. The two Raman thresholds for the RFA are close to the cases when only the temporal stable pump laser is applied in the RFA. Further combining the results shown in Fig. 8 and Fig. 9, it might be concluded that the temporal stability of the initial inserted pump laser would have significant impacts on the Raman threshold for RFAs, while the impact of the temporal stability of the initial inserted seed laser is slight. Notably, through applying the temporal stable laser as the initial inserted pump laser, both the IM FWM and IM FWM interactions in RFAs could be effectively suppressed, and the Raman threshold for RFAs could be increased by over twice.

The impacts of the intensity fluctuations in the pump and the seed lasers on RFAs could be understood by the noise transfer properties. In high-power RFAs, most of the intensity fluctuations in the initial inserted pump laser would be transferred into the output signal laser while the intensity fluctuations in the initial inserted seed laser could be filtered during amplification [23]. Accordingly, when applying temporal stable laser as the initial inserted pump laser, the intensity fluctuations in both *LP0* and *LS0* along the fiber would be much weaker than the case when there exists strong intensity fluctuations in the initial inserted pump laser. Consequently, the corresponding effective Raman gain would be smaller and the strength of FWM interactions would be much weaker, which finally leads to the enhancement of the Raman threshold for RFAs.

## 5. Conclusions

In conclusion, we derive the coupled amplitude equations which are capable of analyzing the evolutions of different spectral components in different transverse modes for RFAs. Both the effects of the FM FWM and IM FWM on RFAs are demonstrated through numerical simulations. The simulation results reveal that both the FM FWM and IM FWM could lead to significant decline of the Raman threshold for RFAs. Specifically, effective FM FWM interactions could occur despite that the corresponding wave-vector mismatch is rather big. And effective IM FWM interactions could occur with additional generation of the first order Raman Stokes light in the HOM.
To propose the possible strategies to suppress the FWM effects, we also investigate the effects of the intensity fluctuations in the initial inserted pump and seed lasers on RFAs. Due to the diversity of the noise transfer properties, the temporal stability of the initial inserted pump laser would have significant impacts on the Raman threshold for RFAs, while the impact of the temporal stability of the initial inserted seed laser is slight. Notably, through applying the temporal stable laser as the initial inserted pump laser, both the FM FWM and IM FWM interactions in RFAs could be effectively suppressed, and the Raman threshold for RFAs could be increased by over twice. We believe that the theoretical model presented could also be applied for comprehensive investigation of RFAs and the simulation results could provide a well reference for further scaling the output powers of high-power RFAs.